\documentstyle[aps,prl,psfig,twocolumn,epsf]{revtex}
\begin{document}
\draft
\twocolumn[\hsize\textwidth\columnwidth\hsize\csname @twocolumnfalse\endcsname
\title{
Finite Density of States in a Mixed State of $d_{x^2-y^2}+id_{xy}$ 
Superconductor
}
\author{ W. Mao$^{(1)}$ and  A.V. Balatsky$^{(2)}$   }
\date{\today}

\address{$^{(1)}$Department of Physics, Boston College, Chestnut Hill, MA
02167, USA
}
\address{$^{(2)}$ T-Div and MST-Div, Los Alamos National Laboratory, Los Alamos, 
NM 
87545, 
USA}

\maketitle
\widetext
\begin{abstract}
\leftskip 2cm
\rightskip 2cm
We have calculated the density of states of quasiparticles in a
$d_{x^2-y^2}+id_{xy} $ superconductor, and show that in the mixed state
the quasiparticle spectrum remains gapless because of the Doppler shift
by superflow.  It was found that if the $d_{xy}$ order gap
$\Delta_1\propto \sqrt{H}$ as suggested by  experiments, then thermal
conductivity $\kappa \propto \sqrt{H}$ in accord with experimental data
at lowest temperatures. This is an appended version of the paper
published in Phys. Rev. {\bf B 59}, 6024, (1999). We now also discuss
the disorder effects and analyze the H log H crossover at small fields.
We argue that H log H regime is present  and disorder effect is
dominant as the field-induced seconary gap is small at small fields.
\end{abstract}
%\pacs{PACS numbers: 70.20.Pa, 73.20.Dx, 72.17.+a}
 \vskip2pc]
\narrowtext
Based on the experiments recently carried out by several 
groups\cite{Krishana,Aubin} , it has 
been found
that the longitudinal thermal conductivity $\kappa_{xx}$ of unconventional 
superconductor, such as $Bi_2Sr_2CaCu_2O_{8+\delta}$,
%and $YBa_2Cu_3O_{6+x}$
 displays
quite strange behavior in the mixed state. With temperature from 5K
to 20K, it decreases as the applied magnetic field along c-axis increases. At 
some critical field value $H_k(T)$, thermal conductivity becomes insensitive to 
the 
increase 
of magnetic 
field and develops a plateau. In order to explain this phenomena, it has been 
proposed that  
the $d_{x^2-y^2}
$ order parameter is suppressed at $H_k$ due to the opening of a $d_{x^2-y^2}
+id_{xy}$ gap on the whole Fermi surface, thus suppressing quasiparticle 
contribution to
 thermal transport.

  On the other hand, more recent  measurements at lowest temperatures 
\cite{Aubin}, 
say 0.1K,  show thermal conductivity
increases as field rises $\kappa_{xx} \propto \sqrt{H}$, as well as a 
substantial 
history 
dependence of the $\kappa$
in the mixed phase.  These results clearly point towards the relevance of the 
vortex lattice 
and possible
disorder and hysteresis in the vortex lattice for the thermal transport in the  
mixed state. 

From these measurements one can conclude that the spectrum of quasiparticles has 
a 
finite 
density of states $N(E = 0, H)$ at lowest energies.  While, we argue, based on 
the 
available 
data for $\kappa$ at lowest temperature,
 one still can not rule out the transition into fully gapped state, such as 
$d_{x^2-y^2} + i 
d_{xy}$.   

Here we present  the simple model calculation which indeed shows that for fully
gapped state one can still get $\kappa\propto\sqrt{H} $ if $\Delta_1\propto
\sqrt{H} $ as proposed by Laughlin\cite{Laughlin}, and   can reconcile the
fully gapped spectrum of quasiparticle in $d_{x^2-y^2}+id_{xy}$ state and the
experimental data at lowest temperature by Aubin et al\cite{Aubin}.

We repeated Volovik's calculation\cite{Volovik}, in which the spectrum of 
quasiparticles
is modified because of the Doppler shift from superflow in the mixed
state. Volovik found that with the pure $d_{x^2-y^2}$ order parameter,
 the density of
states of the delocalized quasiparticles 
has  relation $N(E=0,H)\propto\sqrt{H}$.
Here we consider $d_{x^2-y^2}+id_{xy}$  order parameter, 
thus 
the quasiparticle spectrum is fully gapped, 
$E({\bf{k}})=\sqrt{\epsilon^2({\bf{k}})+|{\Delta ({\bf{k}})}|^2}
$, where $\Delta ({\bf{k}})=\Delta_0({\bf{k}}) +i\Delta_1({\bf{k}})$ ,
$\Delta_0({\bf{k}})$ represents for the $d_{x^2-y^2}$ order parameter, and 
$\Delta_1({\bf{k}})$ for $d_{xy}$. Below we assume that there is some average 
value 
of the 
gap $\Delta_1({\bf{k}})$ in the vortex lattice  unit cell and ignore its 
position 
dependence. The 
spatial variations of $\Delta_1({\bf{k}})$ albeit substantial, will not change  
qualitatively 
 results presented below.

The DOS for delocalized quasiparticles at Fermi surface can be calculated from
following equation, taking $T=0$,
\begin{equation}
N_{deloc}(0)=2\int \frac{d^3k}{(2\pi)^3} \int d^2{\bf{r}} \delta [E({\bf{k,r}})+
m_e {{\bf{ v}}_F \cdot {\bf{v}}_s}],
\end{equation}
where ${\bf{v}}_s=(\hbar/2m_e)(\widehat{\phi}/r)$.
The main contribution of the DOS comes from the vicinity of the $d_{x^2-y^2}$
gap nodes, the gap function has the general form around the node $ \Delta_0
({\bf{k}})\approx {\bf{\vec\gamma}}(k_z)\cdot [{\bf{k}}-{\bf{k}}_n(k_z)]$,
where $\vec\gamma(k_z) || {\bf{e_z}}\times {\bf{k}}_n.$
It is convenient to introduce  new momentum variables $
\widetilde{k}_x=({\bf{k}}-{\bf{k}}_n)\cdot \widehat{k}_{n},
\widetilde{k}_y=({\bf{k}}-{\bf{k}}_n)\cdot \widehat{\gamma}_n,
$ where $\widehat{k}_{n}$ is the unit vector in the direction of the gap
 nodes in the $a-b$ plane, $\widehat{\gamma}_n$ the unit vector along the
direction of $\vec\gamma$.
Then we can write Eq.(1) as
\begin{eqnarray}
 & & N_{deloc}(0)=4\times \frac{1}{4\pi^3} \int d\widetilde{k}_y 
dk_z \frac{d\epsilon}{v(k_z)}
 d^2{\bf{r}} \nonumber \\
 & & \delta[
\sqrt{\epsilon^2+{\widetilde{k}_y}^2\gamma^2(k_z)+{\Delta_1}^2}+
m_e {\bf{v}}_F ({\bf{k}}_n)\cdot
{ \bf{v}}_s
],
\end{eqnarray}
and the integral can be simplified as
\begin{eqnarray}
 & & N_{deloc}(0)= 
\nonumber \\
 & & \frac{2}{\pi^2v_F\gamma}\int d k_z {\int}d\theta
 {\int_\xi}^{R_H}rdr
{\int_{0}}^{{k_c}^2}
d x \delta(\sqrt{x+{\Delta_1}^2}+{k_F}{v_s} cos\theta).
\end{eqnarray}
Substituting $\sqrt{x+{\Delta_1}^2}$ by the variable $y$, we can further 
write above equation as
\begin{equation}
N(0)=\frac{2}{\pi^2 v_F\gamma}{\int_{\Delta_1}}^{\infty} d y 
\int d k_z {\int}d\theta {\int_\xi}^Rrdr y \delta(y+k_F v_s cos\theta).
\end{equation}
In order to find the analytical form we first
express the delta function as an integral form,
$
\delta(y+k_F v_s cos\theta)=\int_{-\infty}^{\infty} 
d \lambda \exp{(i \lambda y - i \lambda k_F v_s
cos\theta)},
$
and integrate over  $\theta$, then over  $\lambda$. We find
  from Eq.(4),
\begin{eqnarray}
 & & N(0)=\int \frac{2dk_z}{\pi v_F\gamma}
{\int_\xi}^{R_H} r dr \sqrt{{k_F}^2 {v_s}^2 -{\Delta_1}^2}
\Theta({k_F}^2 {v_s}^2 -{\Delta_1}^2) \nonumber \\
 & & =\int d k_z \frac{2\Delta_1}{\pi v_F\gamma}
{\int_{\xi}}^{Min(r_0,R_H)} dr \sqrt{{r_0}^2-r^2}
\nonumber \\
 & & \approx \int d k_z \frac{\Delta_1}{\pi v_F\gamma}
(R_H \sqrt{{r_0}^2-{R_H}^2} + r_0^2 arcsin{\frac{{R_H}}{r_0}})
\end{eqnarray}
where $r_0=\frac{1}{2}\xi \widetilde{\Delta}^{-1}$, 
$\widetilde{\Delta}={\Delta_1}/{\Delta_0}$, 
$R_H$ is the intervortex
distance, and $\xi$ is the coherent length for pairing. We already 
know that $R_H \approx
\xi\sqrt{H_{c2}/H}$,
and assume $\Delta_1$ much smaller
compared with $\Delta_0$, hence $Min(r_0, R_H)=R_H$. After making Taylor 
expansion in terms of 
$R_H / r_0$,
%$\widetilde{\Delta}$,
we can get the following relation averaged over the vortices
\begin{equation}
 N(0)\approx K N_F \widetilde{\Delta} (\frac{ r_0}{R_H} -\frac{1}{3}\frac{{R_H}}
 {{r_0}}) \sim {\sqrt{H}} - \frac{\Delta^2_1}{\sqrt{H}}.
\end{equation}
Here the factor K is on the order of unity.
We found that the spectrum remains gapless. The origin of the gapless behavior 
is 
the Doppler shift of quasiparticle states
$ E - {\bf k v}_s$, which is position dependent. There are regions where this 
shift
 is larger than 
minimal gap in the spectrum $\Delta_1$, thus leading to the  finite DOS. This
 point was also emphasized      by  Hirschfeld and  W{\"o}lfle, who
   showed that for finite
 $id'$ gap the density of states will  be finite \cite{HW}.  We note that 
regardless of the power with which $\Delta_1 \sim H^{\alpha}$ opens up in the 
field, as long as $\alpha \geq 1/2$, the leading term in DOS will always be 
$\sqrt{H}$ at small fields.

If one uses the form of $\Delta_1= 
\hbar v\sqrt{2eH/(\hbar c)}$ used by Laughlin\cite{Laughlin} ($v$ is the 
average 
quasiparticle velocity in  the
$d$-wave node) 
to explain 
Krishana's data, we 
find that for the DOS of the fully gapped state
\begin{eqnarray}
& &N(E=0,H)=KN_F(\sqrt{\frac{H}{H_{c2}}}-\frac{4}{3}\frac{\hbar e v^2}{c 
\Delta_0 ^2}
\sqrt{H_{c2} H}) \nonumber\\
& &=KN_F\sqrt{\frac{H}{H_{c2}}}(1-\frac{4\Delta_0}{3E_F}),
\end{eqnarray}
where $E_F$ is the Fermi energy. Typically $\Delta_0/E_F \sim 10^{-1}$ and the 
effect of induced gap is to lower the
 $\sqrt{H}$ prefactor in $N(0)$  by about 20\%. 
The  DOS with $N(E=0,H) \sim \sqrt{H}$ was previously thought to be 
characteristic 
of 
``pure" $d_{x^2-y^2}$
state\cite{Volovik}.

Experimental result by Aubin et al.\cite{Aubin} showed that the
thermal conductivity at lowest temperature with $\kappa(H)\propto\sqrt{H}$
 is  consistent 
with quasiparticle transport and ``Volovik effect'' in the density of states. 
They also pointed out that even at  lowest temperature data are  consistent with 
the 
gapped $d_{x^2-y^2}+id_{xy}$ phase provided $\Delta_1$ is small enough.

 \begin{figure}
\epsfxsize=3.9in
\centerline{\epsfxsize=8cm \epsfysize=6cm \epsfbox{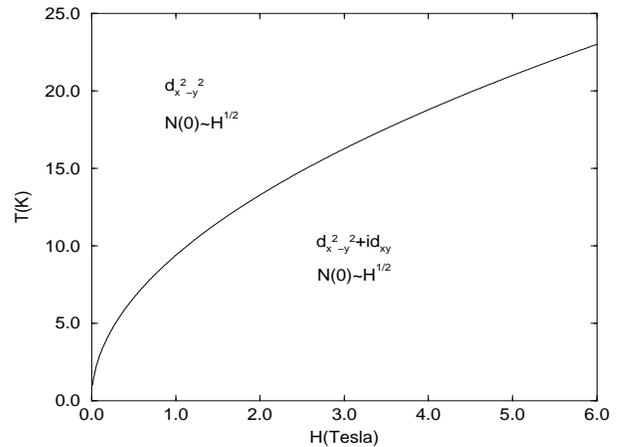}}
\caption[]{
Two regions of the phase diagram are shown. The line is the the boundary between 
gapped and gapless states, given by
 $k_BT_c = 0.52 \hbar v\sqrt{{2eH\over{\hbar c}}}$. In both phases the density
of states $N(0,H)\propto\sqrt{H}$ due to $\Delta_{xy}\propto\sqrt{H}$.  }
\end{figure}

Fig.1 demonstrates that the superconductor has two phases\cite{Laughlin} as 
temperature and
applied magnetic field change, one with ``pure" $d_{x^2-y^2}$ order parameter,
the other with $d_{x^2-y^2}+id_{xy}$ order parameter. According to our 
calculation,
in both phases, the DOS of quasiparticle at $E=0$ is proportional to
square root of applied field.

We also note the implication of our calculation for the specific heat.
Recent measurements of the specific heat $C(T, H)$ in
 the mixed 
state of YBCO superconductors
by Moler et.al. \cite{Moler} indicate that the low temperature density of states
 indeed scales as $N(E=0,H) \sim \sqrt{H}$. This observation however, 
 as we argue, is not inconsistent with the $d+id'$ gap in the  mixed state.

In the paper by Volovik\cite{Volovik}, it was mentioned that in the mixed state
of the d-wave superconductor, the conventional s-wave pairing will be
 generated in the core of vortex. This might be another possibility to 
obtain fully gapped
quasiparticle spectrum. However, it will not affect our result which
comes from the region outside the core.

The main result of this work is to show that even in the presence of fully 
gapped spectrum, the superflow modifies quasiparticle
spectrum and 
keeps  DOS  gapless in the mixed state. Approximating  
thermal conductivity
as proportional to the DOS at $E=0$ at lowest temperature, 
and assuming that the magnitude of ${\Delta}_1$ proportional
 to $\sqrt{H}$ in accordance
with the work by Laughlin, one can still be  consistent with 
experimental data at lowest temperature. It shows that we cannot rule out
 the possibility that a second superconducting phase 
appears in the magnetic field at low temperatures.

Note Added (March, 1999):
In the paper by C. Kubert and P.J. Hirschfeld\cite{Kubert}, the density of 
states was calculated in a 
$d_{x^2-y^2} $ superconductor in the dirty limit, and they found a $H \ln H$ 
behavior at the lowest field instead of $\sqrt{H}$. Here we also calculated the 
density of states with the induced gap $\Delta_1$ in the presence of impurity 
scattering, and found that the appearance of the induced   gap $\Delta_1$ does 
not change the $H log H$ behavior in the lowest field. The details of 
calculation are described in the appendix.

We are  grateful  to K. Bedell, M. Franz, R. Movshovich, M. Salkola, J. Sauls 
and  P. 
W{\"o}lfle
for useful  discussions. This work was supported by DoE. W.M.   acknowledges the 
LANL  support  for the visit.

\appendix
\subsection{Appendix A}
We assume the impurity gives an imaginary term $\gamma_0$ in the self energy of 
one particle Green function. Thus the density of states can be calculated by 
following equation:
\begin{eqnarray}
& & N(E=0,H)=\frac{1}{\pi R_H^2} \int d^3 {\bf k} \int d^2 {\bf r} Im(G)
  \nonumber\\
& & =\frac{1}{\pi R_H^2}
 \int d^3 {\bf k} \int d^2 {\bf r}
 \frac{\gamma_0}{({\bf v}\cdot{\bf k} 
+\sqrt{\epsilon_k^2+|\Delta|^2})^2+\gamma_0^2}
\nonumber\\
& & =\frac{1}{ R_H^2 \gamma v_F }
\int rdr \int d\theta \int dk_z \int dx 
\frac{\gamma_0}{({\bf v}\cdot{\bf k_F} +\sqrt{x+\Delta_1^2})^2+\gamma_0^2}
\nonumber\\
& & =\frac{\gamma_0}{ R_H^2 \gamma v_F }
\int rdr \int d\theta \int dk_z \int_{\Delta_1}^Y \frac{ydy}{({\bf v}\cdot {\bf 
k}+y)^2+\gamma_0^2} \nonumber\\
& & = \frac{\gamma_0}{\gamma v_F} \int d k_z \int_{\frac{\xi}{R_H}}^1 xdx \int
d\theta \{\frac{E_H}{\gamma_0} \frac{cos\theta}{x}[arctan(\frac{\frac{E_H 
cos\theta}{x}+
\Delta_1}{\gamma_0}) \nonumber\\
& & -\frac{\pi}{2}] 
 +\frac{1}{2}
ln \frac{(\frac{E_H cos\theta}{x}+
Y)^2+\gamma_0^2}
{(\frac{E_H cos\theta}{x}+
\Delta_1)^2+\gamma_0^2}\},\nonumber
\end{eqnarray} 
where $E_H=\sqrt{H/H_{c2}}\Delta_0$, $Y$ is an upper  cutoff of 
$\sqrt{\epsilon_k^2+|\Delta|^2}$. It should be noticed that $\gamma_0$ is the 
imaginary self energy term from impurity, while $\gamma$ is the gradient of 
$\Delta_0$ with respect to $\widetilde{k}_y$

In above equation, we can get the $H log H$ behavior when $\Delta_1$ equals to 
zero. We calculated the integral numerically, including both the effect of 
induced $d_{xy}$ gap and impurity, and show them in Fig.2. In our calculation we 
used the result, $\gamma=0.61\sqrt{\Gamma\Delta_0}$, of Kubert and 
Hirschfeld\cite{Kubert}, with $\Gamma=n_i/ \pi N_F$. It is found  that the 
presence of an induced $d_{xy}$ gap still cannot change the behavior of the 
density of states as a function of  applied field, however it decreases the 
density of states a little bit.  

 \begin{figure}
\epsfxsize=3.9in
\centerline{\epsfxsize=8cm \epsfysize=6cm \epsfbox{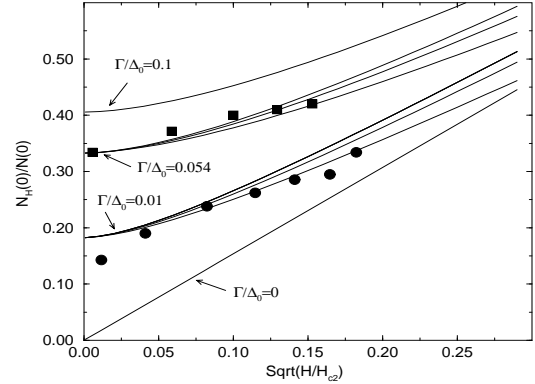}}
\caption[]{
Normalized density of states vs $(H/H_{c2})^{1/2}$. The three line groups for 
each  data 
sets correspond to different strength of $\Delta_1 = x \Delta_0 
(H/H_{c2})^{1/2}$ with $x = 0, 0.5, 0.8$ from top line to bottom. The presence 
of the {\em field induced} $\Delta_1 \propto H^{1/2}$ does not change the 
$N_H(0)$ behavior at lowest fields as the effect of disorder and $\Delta_0$ 
dominates this region. Hence for dirty $d_{x^2-y^2}+id_{xy}$ we expect the same 
$H^{1/2} \rightarrow H \ln H$ crossover, as discussed before \cite{Kubert}. 
The data are from references [6] and [8].} 

\end{figure}

\end{document}